\documentstyle[prl,aps]{revtex}
\begin{document}
\draft
\title{
Polaron Formation in the Three-Band  Peierls-Hubbard Model for
Cuprate Superconductors cond-mat/9401025 }
\author{J. Lorenzana}
\address{
Laboratory of Applied and Solid State Physics, Materials Science Centre,\\
University of Groningen, Nijenborgh 4, 9747 AG Groningen, The Netherlands
}
\author{A. Dobry\cite{perad}}
\address{
Groupe de Physique Statistique, Universit\'{e} de Cergy Pontoise,
47 Avenue des Genottes,95806 Cergy Pontoise Cedex,  France
}
\date{\today}
\maketitle

\begin{abstract}
Exact diagonalization calculations show a continuous
transition from  delocalized to small polaron behavior as a function
of intersite electron-lattice coupling.
A transition, found previously at Hartree-Fock level
[Yonemitsu et al., Phys. Rev. Lett. {\bf 69}, 965 (1992)],
between a magnetic and a non magnetic state does not subsist when
fluctuations are included. Local phonon modes become softer
close to the polaron and by comparison with optical measurements
of doped cuprates we conclude that they are close to the transition
region between polaronic and non-polaronic behavior. The barrier to
adiabatically move a hole vanishes in that region suggesting large mobilities.
\end{abstract}


\narrowtext

The discovery of high Tc superconductors\cite{bed86} has triggered a renewed
interest in strongly correlated systems. More recently the  interplay between
strongly correlated carriers  and lattice effects has produced a lot
of experimental and theoretical attention\cite{bar92}. Recent infrared
 (IR) absorption  experiments show
 that added carriers produce substantial lattice relaxation around
them  showing
up in phonon side bands and typical shake-up bands related to Cu-O
stretching modes\cite{tho93,cal93,fal93}. Similar effects were previously
found in photodoped\cite{kim89,tal90} experiments.

The nature of doping states coupled to
the lattice has been studied in inhomogeneous
Hartree-Fock (HF) with random phase approximation (RPA) fluctuations
calculations\cite{yon92,yon93,lor92}. This approach predict the expermentally
observed  doping induced phonon side bands
 as  a consequence of polaron formation, however the
competition between the self-trapping effects and the kinetic energy of the
hole is absent. The same problem arises in a recently introduced
LDA+U scheme\cite{ani92}.

In this work we have performed exact diagonalization and
 HF with RPA fluctuation calculations of the
three-band-Peierls-Hubbard model describing the Cu-O planes.
The exact calculations confirm  the formation of a local-singlet-polaron (LSP)
for large  electron-lattice coupling strength ($\lambda$),
analogous to the lattice polaron state found in HF and LDA+U.
 However the mean field picture is qualitative wrong for small $\lambda$.
Contrary to the HF and LDA+U approaches a minimum electron-lattice
coupling strength ($\lambda_c$) is necessary to produce the polaron
in accordance with the situation in uncorrelated systems\cite{rae86}.
 Furthermore by extending the  previous HF results\cite{yon92}
 and comparing with exact diagonalization we show that a previously  found
 transition between a magnetic and a non magnetic state on Cu is a feature
of the approximations involved and have no counterpart with the exact results.
We show also that as the polaron is formed in hole doped materials
local softening of the phonon modes occur. This explains the
doping induced phonon side bands and by direct comparison with experiments
we can estimate how far from $\lambda_c$ a real material is.
Our result suggest that doped cuprates  are close to the cross-over
region where polaron formation occurs. We find that in that region the
 barrier for adiabatic  motion of a hole vanishes suggesting relatively
large mobilities.

We consider
a  three-band-Peierls-Hubbard model\cite{yon92,yon93,lor92}
 in which for simplicity Cu atoms are kept fix and O ions are allowed to
 move with displacements $u_i$ in the direction of the Cu-O bond.
   Holes have an on-site interaction
$U_d$ on Cu and $U_p$ on O, a Cu-O repulsion $U_{pd}$, on-site energies
on Cu $E_d$, and on O $E_p$, and Cu-O hopping $t$
(O-O hopping is neglected here).  When an O moves in the direction
of one Cu with displacement $|u_i|$ the corresponding Cu-O hopping
changes by $\alpha |u_i|$. Opposite signs apply when the O moves in the
opposite direction. Each Cu-O bond has a spring with force constant
$K$\cite{yon92,yon93,lor92}.
An electron-lattice  coupling constant is defined as
$\lambda=\frac{\alpha^2}{K t}$. Parameters are $E_p-E_d=3, U_d=8, U_p=3,
U_{pd}=1,t_{pd}=1,K=32t_{pd}\AA^{-2}$\cite{yon92,dob93}. Details of the
method are presented in a preliminary report\cite{dob93}.

 The formation of the lattice polaron can be understood by
 repeating Zhang and Rice's\cite{zha88} (ZR) construction in the case in which
the O ions are allowed to relax adiabatically in a symmetric breathing mode
along the Cu-O bonds with displacements
$| u_i|= u$. Following ZR consider first a single
CuO$_4$ cluster
with two holes. The energy  of the local singlet state as a function of $u$ is
 $ E(u)=\epsilon - \beta u + 4 K_p u^2$,
where  $\epsilon= \Delta-8 t^2 (\frac1{\Delta}+\frac1{U_d-\Delta})$
 is the energy of the local singlet in absence of  lattice relaxation.
 $\beta= 16 \alpha t(\frac1{\Delta}+\frac1{U_d-\Delta})$ is
 a renormalized coupling with the lattice.
 $K_p=K- 2\alpha^2(\frac1{\Delta}+\frac1{U_d-\Delta})$ is a renormalized
 spring-force-constant.
 The corresponding phonon frequency of the breathing mode
$\omega_p^2= 2 K_p/M$, with M the
 O mass, is smaller than the one for the ``undoped case'' (only one hole)
$\omega_0^2= 2 K_0/M$, with  $K_0=K-\frac{\alpha^2}{\Delta}$.
  A renormalized electron-lattice coupling
strength can be defined as $\lambda_p=\frac{\beta^2}{K_p t}$. Large $\lambda_p$
implies strong tendency for polaron formation. The expression
 for $\beta$ (note the large  factor in front which comes from
coherence effects), shows that the strong covalency of ZR
wave function favors polaron formation and
competes with $\Delta$ and $U_d$. We can find equilibrium positions
for the lattice displacements $u^0=\frac{\beta}{8K_p}$.
The resulting state is a LSP showing local softening of the lattice
breathing modes. Note that in the electron doped case
(no holes) a hardening is obtain. We show below that IR active modes
in the hole doped case behave in a similar manner in accordance
with previous RPA results\cite{yon92}.
 When the ZR singlet is
allowed to hop, a competition will arise between the self-trapping
and the delocalization effects.

 To further substantiate these findings we have performed a
 Lanczos diagonalization and HF plus RPA
 study of the cluster in the inset of Fig.~\ref{udl}. The
 O ion mass was taken to be infinity (adiabatic approximation),
the O displacements where determined
by iterating self-consistent equations for {\em unconstrained}
 O's displacements.  As we show below this is particularly important
 close to $\lambda_c$. In Fig.~\ref{udl} we show the O
 displacements as a function of $\lambda$ in the two approaches.
The higher branch corresponds to an O neighbor to Cu$_1$ (see inset),
so it represents the equilibrium displacement in the site of the
polaron ($u^0_p$) and the lower
branch correspond to an O neighbor to Cu$_4$ ($u^0_{\infty}$).
 For $\lambda <\lambda_c $
both branches coincide in the exact result and there is no
polaron formation whereas for  $\lambda >\lambda_c $ the particle self-traps
 around Cu$_1$ and the displacement there grows rapidly. The non zero
lattice displacement for $\lambda<\lambda_c $  indicates a
tendency to charge density wave formation (CDW) in which the O's around
Cu$_1$ and Cu$_4$ slightly contract  and is discussed
elsewhere\cite{dob93}. Contrary to the exact results the HF
calculation shows $\lambda_{c,HF}=0$. This is a consequence of the fact
that in this approach the particle is already self-trapped without coupling
to the lattice\cite{yon92,yon93,lor92,lor93a} forming a magnetic polaron state
reminiscent of a local singlet\cite{zha88}. The same is true in LDA+U.
 Hence the competition
between the self-trapping effects and the translational motion of the
hole is absent in these mean field approaches. Note that for large $\lambda$,
when the symmetry is broken in both approachs
HF and the exact result are in good agreement. This is because the
displacements are determined by the one body density matrix which is
optimized in HF. In contrast, as we show below, two body correlations
are badly reproduced.

It was found in HF\cite{yon92}
 that as $\lambda$ approachs a cross-over value $\lambda^*$
a transition occurs to a non-magnetic state in Cu$_1$.
This cross-over is signaled by an increase in the double
occupancy of Cu$_1$ towards the non-magnetic limit
($ \langle n_{\rm\uparrow Cu_1} n_{\rm\downarrow Cu_1} \rangle_{\rm non mag.}
= n_{\rm Cu_1}^2/4$) with $n_{\rm Cu_1}=
\langle n_{\rm\uparrow Cu_1}+n_{\rm\downarrow Cu_1}\rangle $
as shown in Fig.~\ref{ddl} by the triangles.
 In our case $\lambda^*\sim .6$.
We have calculated the evolution with $\lambda$ of the
exact double occupancy at this Cu. The result is shown by the open
squares in Fig.~\ref{ddl}. For $\lambda<\lambda_c$ one should be aware that HF
breaks translational symmetry in a different way to that of  the exact
result, so it is difficult to compare both solutions in detail.
For $\lambda=0$ this can be partially solved by averaging over
the four possible locations of the mean field state\cite{lor92,lor93}
 (large triangle in Fig.~\ref{ddl}).  The result overestimates
the double occupancy.
For $\lambda>\lambda_c$ both the exact and HF results break translational
invariance in the same way. It is clear in that limit that the exact
double occupancy is
well below the non-magnetic limit. We note also that
the sudden increase at $\lambda_c$ does
not reflect in the {\em total}  Cu double occupancy that remains constant
and it is rather an effect due to the rearrangement of the charges.
 To identify the source of the discrepancy we add fluctuations
to the HF double occupancy by calculating the RPA correlation energy and
using the Hellmann-Feynman theorem. Remarkably we find that RPA partially
restores translational invariance so that, contrary to the HF case, there
is little difference for $\lambda\sim 0$ and $\lambda\sim \lambda_c$
 between the double occupancy in Cu$_1$ alone and the spatially averaged
value (see Fig.~\ref{ddl}). The intermediate region is discussed below.
In the polaronic region we show the result at Cu$_1$ whereas in the
delocalized region the spatial average is used (small circles).
We also show the result at Cu$_1$ for $\lambda=0$ (big circle).
  As $\lambda^*$
is approached fluctuations become too big and the RPA breaks down.
This is obvious in Fig.~\ref{ddl} where at $\lambda^*$
the RPA overestimate the correction and a non-physical zero
value of the double occupancy is obtained. If it is calculated on Cu$_1$
alone it becomes negative. In the polaronic region the correction is
overestimated but it is clear from both approaches that the Cu remains
magnetic.

One should note that whereas HF shows a transition between a magnetic
and a non-magnetic state in Cu$_1$ the two solutions in HF {\em plus} RPA
 above and below $\lambda^*$ represent the {\em same} physical situation in
two different ways.
Since the effect of increasing $\lambda$ is to relax the O towards the Cu
we can use $\lambda$ as a measure of the local covalency in the polaron.
Consider the CuO$_4$ cluster with two holes
as before. In the limit in which covalency dominates ($\lambda>\lambda^*$ )
the HF  wave function has the form  $|HF\rangle =\prod_{\sigma}
(\phi_p p^{\dagger}_{\sigma}+\phi_d d^{\dagger}_{\sigma})|0\rangle$
where $p^{\dagger}_{\sigma}=\frac12 \sum_{j=1}^4 p^{\dagger}_{j\sigma}$
and the sum runs over the four O. $p^{\dagger}_{j\sigma}$
 ($d^{\dagger}_{\sigma}$) creates a hole on O (Cu).
 This wave function has spin zero but large
double occupancy on Cu given by  $\phi_d^4$. For an occupancy on Cu of
order one this correspond to $\sim 1/4$ as  is the case in Fig.~\ref{ddl} for
large $\lambda$. RPA introduces correlations and this reduces double occupancy
as is clear in Fig.~\ref{ddl} for large $\lambda$. The net result is
a singlet with small double occupancy which looks very much like the
ZR wave function. On the other extreme when covalency
is not so dominant [$\lambda<\lambda^*$] the HF wave function has broken
symmetry and the single particle orbitals depend on spin, that is
 $|HF\rangle=\prod_{\sigma}
(\phi_{p,\sigma} p^{\dagger}_{\sigma}+\phi_{d,\sigma}
d^{\dagger}_{1\sigma})|0\rangle$. This describes a magnetic state on Cu,
say with up spin, and on O with opposite spin when
$ \phi_{d,\downarrow}^2 < \phi_{d,\uparrow}^2 $ and
$ \phi_{p,\uparrow}^2  < \phi_{p,\downarrow}^2$. The HF wave function
is not a singlet now but RPA corrects for that through  terms
of the form $ J_{\rm Cu-O}\frac12(S^+_dS^-_p+ S^-_dS^+_p)$ present in
the residual interaction. Here $ J_{\rm Cu-O}$ is the Cu-O exchange.
 RPA also reduces further the double occupancy
and the net result is, as before, similar to a ZR wave function.
Now is evident that the transition between these two HF solutions has no
physical meaning but is a consequence of how HF plus RPA reproduce the
physics in two different situations. An analogy can be made
with the problem of a magnetic impurity  in a metal.  Anderson's
classical treatment\cite{and61} shows also a magnetic and a non-magnetic
solution at HF level which can be associated with the magnetic and
non-magnetic state of the Cu. Note the similarity between
Fig.~5 in Ref.~\cite{and61} and Fig.~1(b) in Ref.~\cite{yon92}.
Our result shows the danger of using
mean field alone as a criteria to distinguish between magnetic and
non-magnetic states.

To study how the LSP moves through the lattice
we had parametrized the lattice displacement which interpolates
between the LSP in Cu$_1$ and in Cu$_4$\cite{zho92}. Note that the
transition between nearest neighbors Cu is not allowed in a N\'eel
ordered infinite lattice without producing excitations\cite{aue91a}.
 In Fig.~\ref{edu} we
show the total energy as a function of a collective coordinate
($\delta$). The displacement are set as
$u_i=\frac12(u_p^0+u_{\infty}^0)\pm\delta$ and the signs are chosen
in such a way that for $\lambda >\lambda_c$ and
 $\delta=\frac12(u_p^0-u_{\infty}^0)$ the
configuration of the polaron sitting in Cu$_1$ is obtained.
For $\lambda<\lambda_c $ there is a single minimum which corresponds
to no polaron formation  whereas for $\lambda>\lambda_c$ there are two
minima which correspond to the LSP siting in  Cu$_1$ or in Cu$_4$.
Note that at the transition point the barrier vanishes. Is essential
for this result to consider a {\em full relaxation} of all O in the
cluster. If  only the O around Cu$_1$ are relaxed, the barrier
never vanishes and the transition becomes first order like\cite{dob93}.
This is qualitatively different from a previous Holstein-Hubbard
study\cite{zho92}. There coupling with O c-axis displacements was considered
and the transition between the polaronic and non-polaronic state was
found to be always first order like.

Many experiments show local softening  of IR modes upon hole
doping\cite{tho93,kim89,tal90} or hardening for electron doping\cite{cal93}.
 To study this effect we had made a frozen
phonon calculation by moving in phase the two O at the right and left
of Cu$_1$ and computed the corresponding phonon frequency ($\omega_p$) and
compared with the result in the undoped case ($\omega_0$)
( inset of Fig.~\ref{edu})\cite{note0}.
 The local softening effect is particularly well resolved in
photodoped La$_2$CuO$_4$\cite{kim89,tal90} where  the 708 cm$^{-1}$
Cu-O stretching mode gets bleached and a new band grows at 640 cm$^{-1}$.
This corresponds to  $\omega_0^2/\omega_p^2\sim1.22 $) and a
 $\lambda \sim10\%$ higher than $\lambda_c$\cite{note1}.
This result, and the fact that effective
masses in real materials are  not very large, point to a situation
where the real $\lambda$ in the cuprates is close to $\lambda_c$.
The physics close to that point is highly non-trivial for a finite
value of the ion mass\cite{rae86,zhe88}.
 The transition is not any more sharp but
is washed out by quantum fluctuations and $\lambda_c$ becomes a cross over
value. The renormalization of the
effective mass of quasiparticles is moderate and not exponentially large
like in the large $\lambda$ limit.
An appealing scenario is that for small doping polarons are formed
with $\lambda$ close to $\lambda_c$. This will result in a small
 activation energy for thermal
 diffusion of the hole as is observed in lightly doped
cuprates\cite{cal93,fal93}.
For higher doping, polaron formation is inhibited due to phase space
restrictions and a more complicated object occurs sharing characteristics of
the self-trapped and non-self-trapped state and hence more metallic behavior.

Very recently an infrared  band at .1 eV in both hole\cite{fal93}
and electron hole\cite{cal93} doped compounds has been identified
 as arising from  shake-up process of a polaron phonon cloud.
In the electron doped case\cite{cal93} it was  possible
to separate the  phonon contribution which appears as higher harmonics
of the Cu-O stretching modes considered here, indicating the relevance
of these modes in forming the polaron. Similar
structures are present in the hole doped data\cite{fal93,calper}.

We believe  that these effects can be generic to a wide class of
materials and provide a framework for classification.
For example low mobilities in compounds like
La$_{2-x}$Sr$_x$NiO$_{4+\delta}$\cite{tra93,che93,ani92}.
 can be understood as arising from a value of
 $\lambda>>\lambda_c$.  In the former the self trapping  probably inhibits
the softening of the MIR band\cite{xia93} observed in superconducting
compounds\cite{uch91} for which a physical picture was  recently
presented\cite{lor92,lor93a}. On the other hand we suggest
that for La$_{2-x}$Sr$_x$CuO$_4$ probably $\lambda\sim \lambda_c$ where
interesting physics can arise.

In summary, we have found a continuous transition between
 a LSP polaron and a delocalized state in a three-band-Peierls-Hubbard model.
We showed that there is no quenching of magnetic moments as found in
previous approachs and by comparison with optical measurements we suggest
that cuprates are close to the transition region between polaronic and
non-polaronic behavior. This leads to a rich
 phenomenology where lattice anharmonicities coexist with presumably
high carrier mobilities.

This investigation was partially
supported by the Netherlands Foundation for
Fundamental Research on Matter (FOM) with financial support from the
 Netherlands Organization for the Advance of Pure Research (NWO).
The authors are supported in this work by  postdoctoral fellowships
granted by the  Commission of the European Communities.
 A.D. is also indebted to Fundaci\'on Antorchas of Argentina for
traveling support. We would like to thank J. Riera for kindly
giving us a version of his Lanczsos program and for useful discussions
and to G. Sawatzky, D. van der Marel, P. Calvani, M. Capizzi and S. Lupi
for many insights.


\begin{references}
\bibitem{note0}
This modes are expected to mix very
 little with other modes and can be considered eigenmodes.
\bibitem{note1}
We prefer not to extract an absolute value of $\lambda$ since we
expect that our estimate of $\lambda_c$ should decrease in larger
systems\cite{dob93}. On the other hand the magnitude of the softening
as a function of the relative distance from the transition
$(\lambda-\lambda_c)/\lambda_c$ should be much size
 independent.
\end{references}

\begin{figure}
\caption{
Equilibrium positions for one oxygen ion surrounding Cu$_1$ (higer branches)
and for another surrounding Cu$_4$ (lower branches) versus
$\lambda$.  The triangles (squares) are the HF (exact) result.
The inset shows schematically the cluster studied in the presence of
a polaron in Cu$_1$.}  \label{udl}
\end{figure}

\begin{figure}
\caption{
Double occupancy versus $\lambda$. Full small (big)
triangles are the HF result on Cu$_1$ alone (averaged over the four O),
open triangles are the unpolarized
limit on Cu$_1$, open squares are the exact result on Cu$_1$
 and full circles the HF plus RPA result averaged on the four Cu's
($\lambda<\lambda_c$) and on Cu$_1$ alone ($\lambda>\lambda_c$). We also
show the averaged result for $\lambda=0$ (full big circle).}
\label{ddl}
\end{figure}
\begin{figure}
\caption{
 Total energy as a function of the collective coordinate $\delta$
for $\lambda=.7$ (diamonds), 1.05 (crosses) and 1.4 (circles).
The inset shows the ratio $\omega_p^2/\omega_0^2$ versus $\lambda$.
}
\label{edu}
\end{figure}
\end{document}